\documentclass[sigconf]{acmart}
\usepackage{listings}
\usepackage{subfigure}

\usepackage{listings}
\usepackage{xcolor}

\definecolor{codegreen}{rgb}{0,0.6,0}
\definecolor{codegray}{rgb}{0.5,0.5,0.5}
\definecolor{codepurple}{rgb}{0.58,0,0.82}
\definecolor{backcolour}{rgb}{0.95,0.95,0.92}
\lstdefinestyle{mystyle}{
    backgroundcolor=\color{backcolour},   
    commentstyle=\color{codegreen},
    keywordstyle=\color{magenta},
    numberstyle=\tiny\color{codegray},
    stringstyle=\color{codepurple},
    basicstyle=\ttfamily\footnotesize,
    breakatwhitespace=false,         
    breaklines=true,                 
    captionpos=b,                    
    keepspaces=true,                 
    numbers=left,                    
    numbersep=5pt,                  
    showspaces=false,                
    showstringspaces=false,
    showtabs=false,                  
    tabsize=2
}

\AtBeginDocument{%
  \providecommand\BibTeX{{%
    \normalfont B\kern-0.5em{\scshape i\kern-0.25em b}\kern-0.8em\TeX}}}

\setcopyright{acmcopyright}
\copyrightyear{2018}
\acmYear{2018}
\acmDOI{10.1145/1122445.1122456}

\copyrightyear{2021}
\acmYear{2021}
\setcopyright{rightsretained}
\acmConference[RecSysChallenge 2021]{RecSysChallenge '21: Proceedings of the Recommender Systems Challenge 2021}{October 1, 2021}{Amsterdam, Netherlands}
\acmBooktitle{RecSysChallenge '21: Proceedings of the Recommender Systems Challenge 2021 (RecSysChallenge 2021), October 1, 2021, Amsterdam, Netherlands}\acmDOI{10.1145/3487572.3487599}
\acmISBN{978-1-4503-8693-7/21/10}

\begin{document}

\title[Synerise at RecSys 2021]{Synerise at RecSys 2021: Twitter user engagement prediction with a fast neural model}


\author{Michał Daniluk}
\affiliation{%
  \institution{Synerise}
  \country{Poland}
}
\affiliation{%
  \institution{Warsaw University of Technology}
  \country{Poland}
}
\email{michal.daniluk@synerise.com}

\author{Jacek Dąbrowski}
\affiliation{%
  \institution{Synerise}
  \country{Poland}
}
\email{jack.dabrowski@synerise.com}

\author{Barbara Rychalska}
\affiliation{%
  \institution{Synerise}
  \country{Poland}
}
\affiliation{%
  \institution{Warsaw University of Technology}
  \country{Poland}
}
\email{barbara.rychalska@synerise.com}

\author{Konrad Gołuchowski}
\affiliation{%
  \institution{Synerise}
  \country{Poland}
}
\email{konrad.goluchowski@synerise.com}

\renewcommand{\shortauthors}{Daniluk et al.}

\begin{abstract}
In this paper we present our 2nd place solution to ACM RecSys 2021 Challenge organized by Twitter. The challenge aims to predict user engagement for a set of tweets, offering an exceptionally large data set of 1 billion data points sampled from over four weeks of real Twitter interactions. 
Each data point contains multiple sources of information, such as tweet text along with engagement features, user features, and tweet features. The challenge brings the problem close to a real production environment by introducing strict latency constraints in the model evaluation phase: the average inference time for single tweet engagement prediction is limited to 6ms on a single CPU core with 64GB memory.
Our proposed model relies on extensive feature engineering performed with methods such as the Efficient Manifold Density Estimator (EMDE) - our previously introduced algorithm based on Locality Sensitive Hashing method, and novel Fourier Feature Encoding, among others. In total, we create numerous features describing user twitter account status and content of a tweet. In order to adhere to the strict latency constraints, the underlying model is a simple residual feed-forward neural network. The system is a variation of our previous methods which proved successful in KDD Cup 2021, WSDM Challenge 2021, and SIGIR eCom Challenge 2020. 
We release the source code at: https://github.com/Synerise/recsys-challenge-2021.
\end{abstract}



\maketitle

\section{Introduction}
Twitter is one of the most popular worldwide social media platforms with over 330 million monthly users and approximately 500 million tweets every day\cite{twitter}. 
Their recommendation system has to meet many challenges such as fast recommendation time and adaptation to rapidly changing hashtags and real-world news. A central concept of Twitter recommendations is fairness: the quality of the recommendations should be equally high regardless of the popularity of authors, tweets, and individual languages. All these challenges are reflected in this year's competition hosted by Twitter - the ACM RecSys 2021 Challenge. The aim of the challenge is to predict whether a given user will react to a given tweet. The competing systems are required to provide fair recommendations and work in a strictly resource-constrained setting, requiring very small model size and fast inference.

Our approach takes the second place on the final leaderboard. The model relies on smart input feature preparation with our Efficient Manifold Density Estimator (EMDE) \cite{emde} and Fourier Feature Encoding (see \S\ref{fourier}) methods (among others), used to represent the tweet text, historical interactions of users and their Twitter account status. The underlying model is a simple feed-forward neural network. Our contributions are the following:

\begin{enumerate}
    \item We adapt the EMDE to a new domain and obtain a successful result. Previously, EMDE has achieved top results in session-based and top-k recommendations \cite{emde}, and various challenges: KDD Cup 2021 (predictions in an academic graph) \cite{kddcup}, WSDM Challenge 2021 (recommendation of travel destinations) \cite{daniluk2021modeling}, and SIGIR eCom Challenge 2020 (multimodal retrieval) \cite{sigir_ecom}. 
    \item We introduce the Fourier Feature Encoding which allows to express continuous features in a numerically stable way without normalization.
    \item We analyze the effectiveness of various input features, such as user cluster assignments, user account features, and tweet text features.
    \item We analyze the fairness concept in detail, taking into account tweet languages and user popularity. We find that our model mostly gives fair recommendations.
    \item We show how our design choices lead to the creation of a very efficient system, with single prediction taking about 4ms on a single CPU without a GPU card.
\end{enumerate}



\section{RecSys 2021 Challenge}
The ACM RecSys 2021 Challenge \cite{DBLP:conf/recsys/AnelliKFBTPL0XH21} focuses on the real-world task of predicting whether a user will engage with a given tweet. The competing models must predict the probability of engagement for each of the four reaction types: \textit{Like}, \textit{Reply}, \textit{Retweet} and \textit{Quote}. 
The released data set \cite{belli20212021} consists of more than 1 billion data points from 28 consecutive days between 4 February 2021 and 4 March 2021. 
The data set has a tabular structure with 20 raw features.
The first three weeks were used for training, while the last week was randomly divided into validation and test sets.
At the beginning of the challenge, the models were evaluated on the public leaderboard that used the validation data set. Two weeks before the end of the competition, the validation targets were released as an extra data source for training and a new test leaderboard was introduced. This changed the competition setting as the validation set is sampled from the same time distribution as the final test set, while the training set is sampled from the past.
It imitates the real production environment, where models can be fine-tuned in real-time to adapt to current trends.

The data set contained both positive and negative examples of different types of engagement between a given tweet and the users. The available features describe two types of users, which we further call \textit{engaging user} and \textit{engaged user} for consistency with Twitter nomenclature. \textit{Engaged} user denotes the creator of the given tweet, while \textit{engaging user} denotes the user who reacts (or chooses not to react) to the tweet. Each tweet is represented as set of tokenized wordpiece ids from multilingual BERT model from 66 different languages such as English, Japanese or Thai. Moreover, the data contains information about hashtags, present media, links domains, number of followers of engaged and engaging users, verification of both user's accounts and information if engaged user follows engaging user. 

This is the second edition of the challenge hosted by Twitter, with the previous RecSys Challenge 2020 being very similar in terms of data structure and targets. Four main changes were introduced in the 2021 edition. First, the challenge brings the problem even closer to Twitter's real recommendation systems by introducing strict latency constraints. Each model needed to be uploaded and evaluated in a separate test environment provided by the organizers, with only 1 CPU core, no GPU, $64$GB RAM size, $20$GB disk space, and a time limit of 24 hours for making all predictions, which gives about 6ms per single tweet prediction. It encouraged teams to develop novel solutions that can be easily applied in the production environment.
Second, the size of training data was increased from one week to three weeks. 
Third, the data density was increased in terms of the graph where users are considered to be nodes and interactions as edges. 
Additionally, recommendation fairness was cast as a vital part of the challenge.

The metrics used for performance evaluation were relative cross entropy (RCE) and the average-precision (AP) for each type of engagement. The fairness concept was included in the metrics by dividing authors into 5 groups according to their popularity on the platfom and evaluating separately for each group. The final score was computed as the average of the scores across each group.


\section{Related Work}
The last year's challenge - the Twitter Recsys Challenge 2020 - was very similar to the 2021 edition, with one of the core differences being that it allowed to build more complex models, due to no constraints on latency and smaller data volume. The winners \cite{10.1145/3415959.3415996} introduced an XGBoost model with extensive feature engineering, encoding features with techniques such as Target Encoding, or difference lag (time difference for datetime features). They found that experimental detection of adequate feature transformations and feature combinations was of chief importance. The runner ups \cite{volkovs2020predicting} included extensive feature engineering as well (467 features total). They introduce a Transformer model for comparing the embedding of current tweet and a collection of historical user tweets, overall creating a mixed architecture of deep learning and gradient boosting trees. The 3rd place team \cite{10.1145/3415959.3415994} aimed to exploit correlations between various target variables with a two-stage approach using LightGBM models. 
Overall, top past challenge solutions often included a simple (usually non-neural) classifier model with the bulk of work put into feature engineering. This is the approach whose merit is further confirmed by our submission in a much more resource-constrained setting.

\section{Approach}
The first step of our approach is to obtain a tweet text representation by fine-tuning a DistilBERT model \cite{sanh2019distilbert} on tweets, and using EMDE \cite{emde} to represent text as a \textit{sketch} (a compressed, fixed-size representation of the tweet meaning). We then extract features that describe tweet content,  the engagement history of engaging and engaged users, and their account status features. Text sketch and features are fed into a simple shallow feed-forward neural network which is trained to predict each type of engagement.
Since the time distribution of validation and training sets is different,  we train the model in two stages: 1) using training set, and then 2) fine-tuning the model on validation set, which was released two weeks before the end of the competition. Note that the training set contains 3 weeks of historical data, but validation set is taken from the same time distribution as final test set.


\begin{figure}[ht]%
\subfigure[A non-overlapping 24 hours sliding window was applied to the training period. Engagements from a single day were used as training targets, while all the rest are used for feature extraction.]{
  \includegraphics[width=82mm]{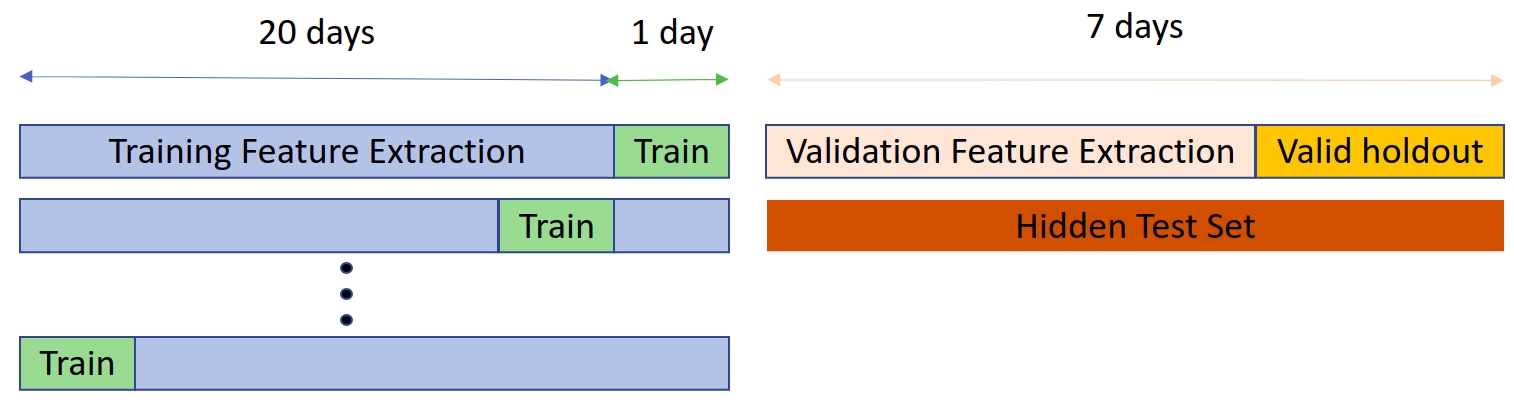}
}
\subfigure[A validation set was randomly divided into 10 parts. Each of them was used as training targets, while the rest 9 parts were used for feature extraction.]{
  \includegraphics[width=82mm]{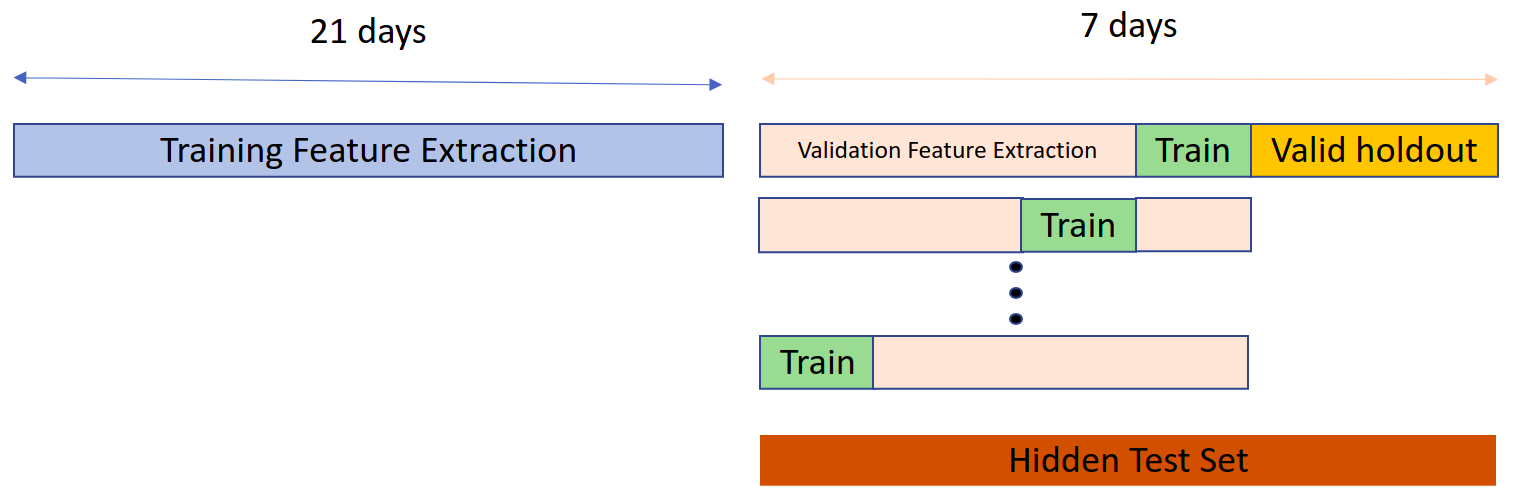}
}
\caption{Data Partitioning. First we train our model using training set, then we fine-tune it using validation set which was sampled from the same time distribution as hidden test set.}
\label{fig:data_partition}
\end{figure}


\subsection{Data Partitioning}
The data partitioning process is depicted in Figure \ref{fig:data_partition}.
Inspired by \cite{volkovs2020predicting}, we apply a non-overlapping 24 hours sliding window to the training period. As a result, all training examples are divided into day-sized chunks based on engagement time for positive examples or tweet creation date for negative samples. Engagements from a single day were used as training targets, while all the rest are used for feature extraction. Note that temporal structure was broken here, because we use features from the future interactions to predict current engagement. This partitioning results in creating 21 training parts that were shuffled before training.
The local evaluation of our model was performed on a 10\% random sample from the provided validation set.

Additionally, we apply similar partition procedure to the validation set whose targets were released by the organizers, but instead of partitioning the data per day, we randomly divided it into 10 parts. Each of them was used as training targets, while the rest 9 parts were used for feature extraction.



\subsection{Feature extraction}
In order to describe users and target tweets, we apply  feature engineering with a particular focus on historical engagements of tweet creator and engaging users. For each data point, we compute the following features:

\begin{itemize}
    \item \textbf{Interactions between engaged and engaging user}. These features summarize historical engagement between the author of a tweet and engaging user. We simply count the number of engagements for each reaction between the engaged and engaging users on historical part of data. 
    
    \item \textbf{Engaged user interactions.} We extract features that summarize the history of interactions of engaged user by counting the number of interactions of each type of engagement which this author received, which describes the total level of engagement they received. 
    
    \item \textbf{Engaging user interactions.} Similarly to the engaged user features, we count the number of each type of reaction which the engaging user gives to any other user.
    In addition, we calculate the number of interactions of engaging user with the tweets that are in the same language as the target one. We also incorporate historical interactions between the engaging user and hashtags from the current tweet.

    \item \textbf{Interactions between engaging user and users similar to the engaged user.}  We count the number of interactions between engaging user and the users which are similar to engaged user. Similar users are detected in the following way: first, for each user $U$, a set of users $F_U$ who follow the user $U$ is created. Then, for each user pair $(A,B)$ we compute their follower similarity score as Jaccard similarity between the follower sets $F_A$ and $F_B$ using an efficient Python set similarity implementation\footnote{\url{https://github.com/ekzhu/SetSimilaritySearch}}. If the similarity score falls above a threshold (selected experimentally), we mark the two users as similar. Clusters of similar users are precomputed at the training stage and reused during inference.
    
    \item \textbf{Interaction with tweets.} Since some tweets from the validation set also appear in the test set, we count the number of interactions with a specific tweet id from the validation set.
    
    \item \textbf{Account status.} We extract features that describe the accounts of a both engaged and engaging users such as: number of followers, number of following users, binary flag indicating if the account of a user is verified,  time since account creation.
    
    \item \textbf{Interaction clusters.} We perform community detection on directed graphs of engaged-engaging user interactions utilizing Leiden algorithm\citep{Traag2019} implemented in \textit{leidenalg}\footnote{\url{https://github.com/vtraag/leidenalg}} package. We calculate Modularity Vertex Partitions with default settings on 4 graphs spanned by all 4 engagement types. For each graph we have 2 features: a 0/1 variable indicating if both users belong to the same partition, and the inverse of partition cardinality (zero when users belong to different partitions). This allows us to capture potentially complex communities of mutual interaction between users, with large communities having lower interaction strength.
    
    \item \textbf{Tweet Content.} Features summarizing the content of a target tweet, e.g: number of hashtags, language of a tweet, the presence of additional content (image, video, gif, links, media, domains), the type of tweet (quote, retweet or top level), time of a tweet.
\end{itemize}

\begin{figure*}[ht]%
\centering
\includegraphics[width=150mm]{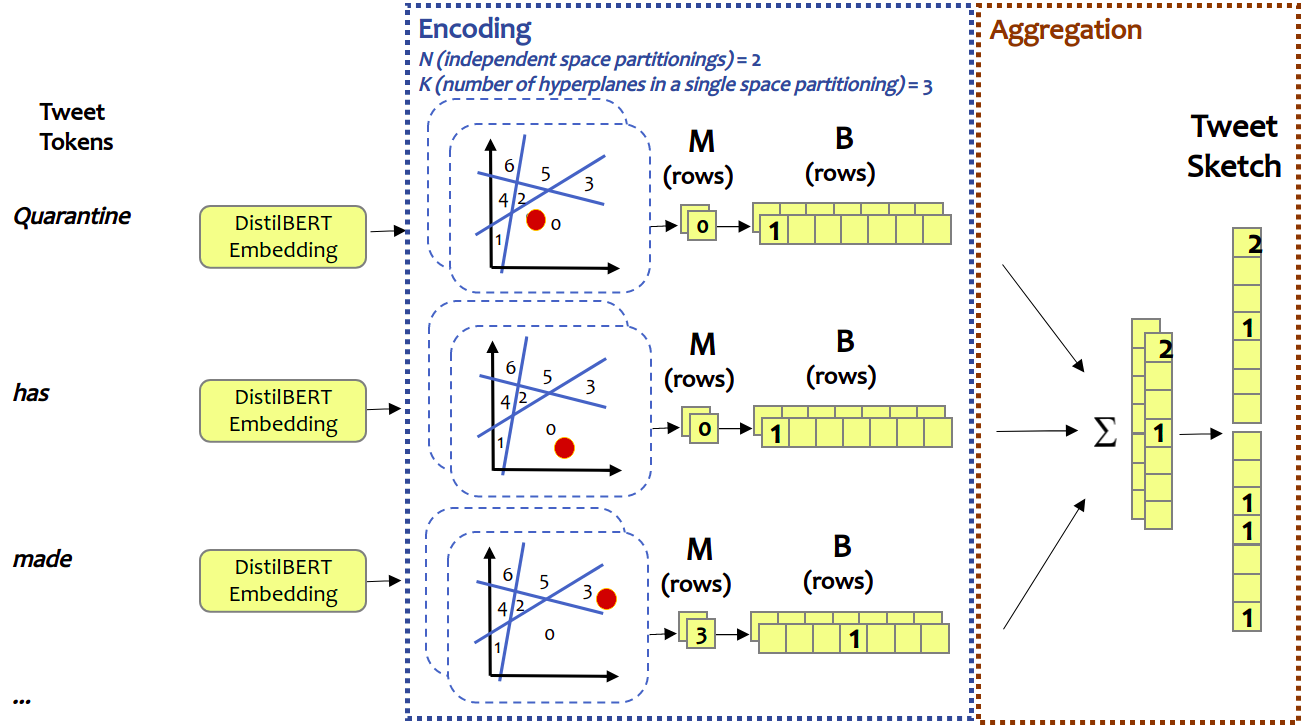}
\caption{EMDE used to represent tweets as token sketches. The matrices $M$ hold ids of regions to which each data point had been assigned after a density-aware LSH space partitioning. The matrices $B$ hold sparsified versions of matrices $M$. Matrices $B$ (representing each word) are summed and normalized to form an aggregate sketch (jointly representing the meaning of all words in a tweet). Aggregate sketches are used directly as input to a downstream model.}
\label{fig:emde-tokens}
\end{figure*}

\subsection{Fourier Feature Encoding}
\label{fourier}
\begin{figure}%
\subfigure[The sine component]{
  \includegraphics[width=40mm]{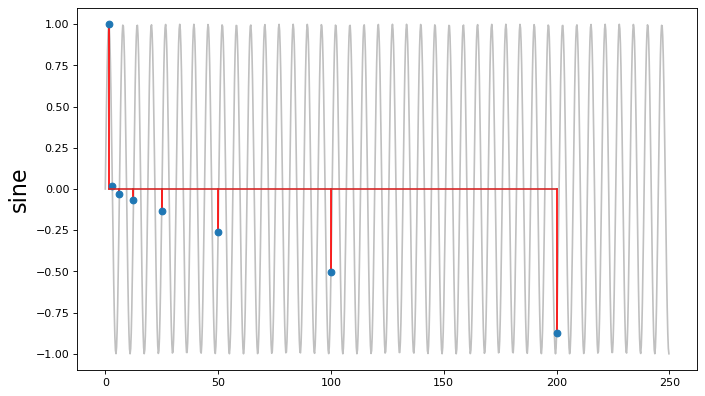}
}
\subfigure[The cosine component]{
  \includegraphics[width=40mm]{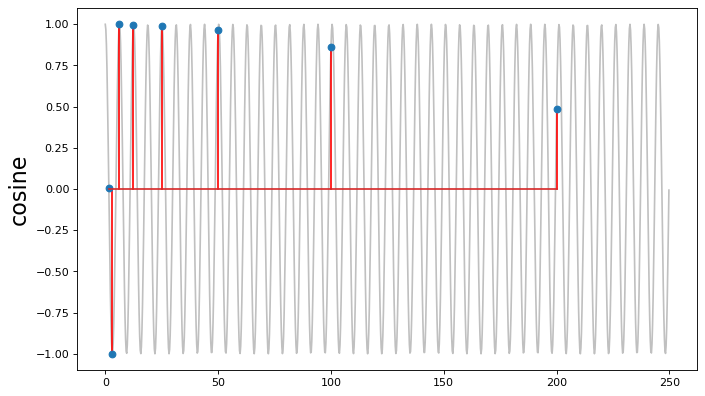}%
}
\caption{Values obtained from the Fourier Encoding for 8 scale levels (powers of 2: $2^1$,...,$2^8$), and input value of 100.}
\label{fig:fourier}
\end{figure}

All numerical features such as number of interactions, number of followers, time since account creations are encoded with our Fourier Feature Encoding, which is a way to sidestep the necessity of explicit normalization of model inputs. Instead of feeding a single input feature, we transform it into a 16-dimensional vector. First, an input numeric value is divided by a few numbers representing  increasing scale levels (in our case, 8 scale levels represented by powers of 2). Then, each of the 8 results is fed to $sin$ and $cos$ functions. The numeric significance of the feature is thus represented on multiple levels, and in a numerically stable way as any number is brought to a small numeric interval defined by trigonometric functions. Fourier Feature Encoding does not usually offer significant performance gains but it facilitates the usage of continuous features. An example of values obtained from the algorithm is shown in Figure \ref{fig:fourier}. Below we present a code snippet of the Fourier Feature Encoding.

\begin{lstlisting}[language=Python]
import numpy as np

def multiscale(x, scales):
    return np.hstack([x.reshape(-1,1)/pow(2., i) for i in scales])
    
def encode_scalar_column(x, scales=[-1, 0, 1, 2, 3, 4, 5, 6]):
    return np.hstack([np.sin(multiscale(x, scales)), 
           np.cos(multiscale(x, scales))])
\end{lstlisting}

\subsection{Tweet representation}
Due to the inference time limit, embedding tweets on the fly with Transformer-based models such as BERT was not feasible. Thus, we utilized a more time-efficient approach representation of tweet contents based on pretrained DistilBERT \cite{DBLP:journals/corr/abs-1910-01108} vectors fed to EMDE \cite{emde}. 
EMDE uses a density-aware manifold partitioning method in order to create meaningful manifold regions. Each region holds samples which are similar according to distance metric expressed within the embedding space which spans the manifold. EMDE can aggregate item sets of arbitrary size into a fixed-side structure which works well with simple feed-forward networks, and much of the computational effort can be done once, before training and inference. 

First, in order to acquire high quality token embeddings, we fine-tune a generic pretrained DistilBERT  multilingual model  \textit{distilbert-base-multilingual-cased} from the Huggingface library \cite{wolf-etal-2020-transformers}. We fine-tuned the DistilBERT model using the training set, then applied the portion of the validation set which is available for training. In order to construct EMDE sketches, we needed to obtain per-token embeddings. We achieved this by embedding all tweets with our DistilBERT model and computing the average per-token embedding. We fit EMDE on a single manifold spanned by all token vectors, irrespective of the language. As a result, we obtain multiple data-aware space partitionings, with similar tokens located in the same regions frequently (per analogy to Locality-Sensitive Hashing \cite{emde}). EMDE sketches have the property of additive compositionality, which allows us to create an aggregate tweet content representation with simple summation of all tweet token sketches. The aggregate per-tweet sketches are finally L2-normalized. The procedure is displayed in Figure \ref{fig:emde-tokens}.

\subsection{Model}
We base our model architecture on our previous design choices which proved appropriate for recommendation with EMDE \cite{emde}, various previous challenges \cite{daniluk2021modeling,sigir_ecom,kddcup}, and which form the backbone of our commercial system. We train a three-layer residual feed-forward neural network with 1500 neurons in each hidden layer, with leaky ReLU activations and batch normalization. The input of the network consists of the following feature vectors, which are simply concatenated:
\begin{enumerate}
    \item Width-wise L2-normalized sketch that represents the text of a tweet.
    \item Numeric features such as number of historical interactions, number of followers etc. which are encoded with our Fourier Feature Encoding.
    \item Categorical features  such as current day, time of tweet and language of tweet that are represented by embedding layers.
\end{enumerate}
The output of the network consists of 4 neurons that represents predictions of like, reply, retweet and quote engagements for a target tweet. The model is trained by optimizing a binary cross entropy loss function for each engagement type.

\section{Experiments}
In this section, we first describe our training setup (§\ref{training}), then we present our final result on the leaderboard (§\ref{results}), and analyze the effectiveness and challenges of our method (§\ref{ablation}).

\subsection{Training}
\label{training}
We train our model on single Tesla V100 16 GB GPU card. Training takes circa 24 hours on one billion training data points. Then, we fine-tune the model on the released validation set, which takes about 60 minutes.
We use AdamW optimizer \citep{loshchilov2017decoupled} with first momentum coefficient of 0.9 and second momentum coefficient of 0.999 \footnote{Standard configuration recommended by \citep{kingma2014adam}} with an initial learning rate of $0.0001$, weight decay of $0.01$ and a mini-batch size of 256 for optimization. The learning rate was linearly decayed.
The final model was trained for 2 epochs on the training set and then fine-tuned for 3 epochs on the validation set.

\subsection{Results}
\label{results}
Table \ref{tab:results} presents results on the final leaderboard. Our team achieves 2nd place in this competition with comparable performance on both AP and RCE metrics.

\begin{table*}[ht]
\centering
\resizebox{\textwidth}{!}{
\begin{tabular}{l|lllllllll}
\textbf{Method}      & \textbf{AP Retweet} & \textbf{RCE Retweet} & \textbf{AP Reply} & \textbf{RCE Reply} & \textbf{AP Like} & \textbf{RCE Like} & \textbf{AP Quote} & \textbf{RCE Quote} & \textbf{Time Taken} \\ \hline
NVIDIA               & 0.4614              & 29.5127              & 0.2649            & 26.6123            & 0.7216           & 23.6124           & 0.0692            & 17.6868            & 23 hours            \\
\textbf{Synerise AI} & 0.4514              & 28.5222              & 0.2559            & 25.7468            & 0.7046           & 22.0994           & 0.0662            & 16.9245            & 18 hours            \\
LAYER6 AI            & 0.4317              & 27.4239              & 0.2490            & 25.3526            & 0.6836           & 19.8578           & 0.0660            & 16.8696            & 13 hours            \\
test\_lightgbm       & 0.4060              & 25.0928              & 0.2118            & 22.6491            & 0.6636           & 17.9193           & 0.0520            & 14.0357            & 5 hours             \\
final1               & 0.3940              & 24.0142              & 0.2077            & 22.1539            & 0.6559           & 16.9609           & 0.0459            & 12.6722            & 19 hours           
\end{tabular}}
\caption{Performance of top five teams on the final challenge leaderborad. Our approach \textbf{Synerise AI} takes 2nd place.}
\label{tab:results}
\end{table*}

\begin{table*}[ht]
\centering
\resizebox{\textwidth}{!}{
\begin{tabular}{l|llllllll}
\textbf{Model inputs}      & \textbf{AP Retweet} & \textbf{RCE Retweet} & \textbf{AP Reply} & \textbf{RCE Reply} & \textbf{AP Like} & \textbf{RCE Like} & \textbf{AP Quote} & \textbf{RCE Quote} \\ \hline
Interactions               & 0.405               & 25.488               & 0.170             & 19.219             & 0.697            & 18.682            & 0.0593            & 14.594             \\
\textbf{+} tweet content features   & 0.427               & 27.382               & 0.221             & 24.614             & 0.7296           & 23.047            & 0.0636            & 16.177             \\
\textbf{+} users account features   & 0.429               & 27.679               & 0.222             & 24.869             & 0.733           & 23.223            & 0.0641            & 16.373             \\
\textbf{+} interaction clusters  & \textbf{0.431}              & \textbf{27.970}                & \textbf{0.227}             & \textbf{25.183}             & \textbf{0.734}            & \textbf{23.676}            & \textbf{0.0663}            & \textbf{16.770}         \\ \hline
\textbf{-} emde                    &    0.424                 &   27.285                   &     0.213              &     24.156               &  0.730                &  23.069                 &   0.0643                &  16.068                                    
\end{tabular}}
\caption{Ablation study results. Features are incrementally added (+) or removed (-) from the model input.}
\label{tab:ablation}
\end{table*}

\subsection{Ablation studies}
\label{ablation}
In order to understand the effects of crucial parts of the training process, we conduct additional experiments. Due to the long training time, we train our models only on $90\%$ of validation set, but use all interactions from the training set as input features. The remaining part of the validation set was used for our final evaluation. 
To ensure a comparable number of parameters of all models, we adjusted the hidden size to have roughly the same total number of model parameters.

The ablation results are summarized in Table \ref{tab:ablation}. In the \textit{Interaction} setting our model takes as input only historical engagement information of target users. By including tweet content features (sketch representation of tweet text; presence of video, image, gif; type of tweet), we observed the average $11.8\%$ increase of the average precision score and $17.4\%$ increase of relative cross entropy  score. The biggest improvement is for \textit{reply} reaction, which may lead to the conclusion that information about the content of the tweet, e.g. information that the author is asking a question, is particularly important for reply engagement.
Adding user account features (number of followers, verification, binary flag indicating if engaging user follows the author of a tweet) improves the AP and RCE scores by $0.5\%$ and $1\%$, respectively. In addition, we verify the impact of features based on user  communities. They increase the average AP score from $0.362$ to $0.365$, and the average RCE score from $23.0$ to $23.4$.

In the next set of experiments, we compare another representation of tweet text. We replace EMDE with an average of all tokens' embeddings learned by our fine-tuned DistilBERT model. Note that embedding the tweets on the fly during inference breaks the latency constraints, so we use precomputed and per-token averaged embeddings. The results are presented in Table \ref{tab:ablation}. 
Elimination of EMDE decreases metrics significantly by $2.8\%$ for AP score, and $2.3\%$ for RCE on average.

\begin{figure}%
\subfigure[Reply]{
  \includegraphics[width=40mm]{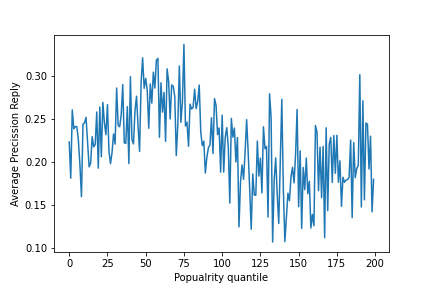}
}
\subfigure[Like]{
  \includegraphics[width=40mm]{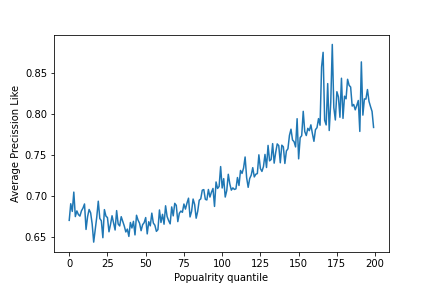}%
}
\subfigure[Retweet]{
  \includegraphics[width=40mm]{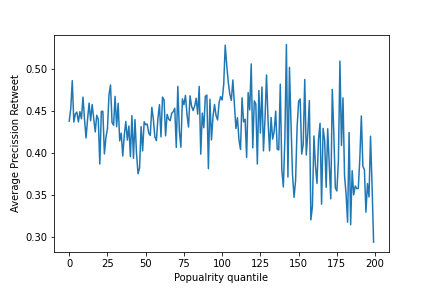}
}
\subfigure[Quote]{
  \includegraphics[width=40mm]{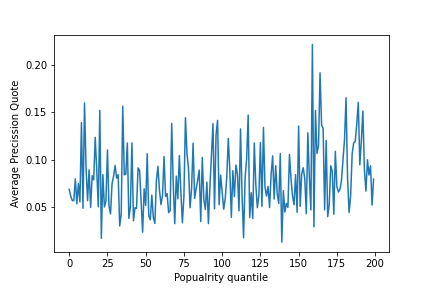}%
}

\caption{Average precision scores for each type of reaction in terms of the popularity of the author of a tweet. Each data point was assigned to one of the 200 quantile groups based on the author's number of followers in test set. The popularity increases with the X-axis.}
\label{fig:fairness}
\end{figure}

We also verify the impact of the fairness concept by dividing the test examples into 200 quantile groups based on the popularity of the author of a tweet (computed as the number of the author’s number of followers). The results are visualised in Figure \ref{fig:fairness}. We can observe that there is no strong trend in the average precision score for \textit{reply}, \textit{retweet} and \textit{quote} reactions. However, the performance of predicting \textit{likes} increases with the popularity of the author.

\begin{figure}[ht]%
\includegraphics[width=70mm]{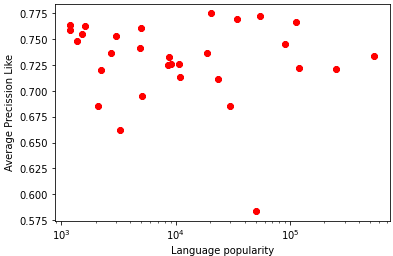}
\caption{Evaluation for different languages of a tweet. The visualization shows the average precision score (y-axis) in terms of the popularity of the language (x-axis), which was computed as number of tweets in particular language.}
\label{fig:languages}
\end{figure}

Additionally, we evaluate our model for different tweet languages. Figure \ref{fig:languages} shows the visualization of average precision score in terms of language popularity. Generally, less popular languages achieve scores similar to most popular languages (with some exceptions). A strong outlier is Thai language, which achieved significantly lower score in spite of being fairly popular. We hypothesize that this is due to the fact that our DistilBERT had been initially trained on 104 languages, however Thai was excluded according to Huggingface documentation\footnote{\url{https://huggingface.co/distilbert-base-multilingual-cased}}.

\section{Summary}
In this paper we present our model which achieves 2nd place in the ACM RecSys Twitter 2021 Challenge. Our model is a 3-layer feed-forward neural network, which ingests tweet text representation encoded with EMDE along with numerical and categorical features that describe users and target tweet. The model is very efficient and adheres to the strict latency constraints within the competition - a single prediction takes about 4ms on a single CPU without a GPU card.

\bibliographystyle{ACM-Reference-Format}
\bibliography{sample-base}

\end{document}